\documentclass[letterpaper, 10 pt, conference]{ieeeconf}  

\IEEEoverridecommandlockouts                              

\usepackage{amsmath} 
\usepackage{amsfonts}
\usepackage{gamealt}

\usepackage{soul} 
\usepackage{cite}
\usepackage{svg}
\usepackage{float}
\usepackage{cuted}

\graphicspath{{fig/}}

\newtheorem{theorem}{Theorem}[section]

\newtheorem{lemma}[theorem]{Lemma}

\title{\LARGE\bf
Rationality and Behavior Feedback in a Model of Vehicle-to-Vehicle Communication
}

\author{Brendan T. Gould and Philip N. Brown
\thanks{Brendan T. Gould and Philip N. Brown are with the department of Computer Science at the University of Colorado Colorado Springs, Colorado Springs, CO, USA {\tt\small \{bgould2,pbrown2\}@uccs.edu}}%
\thanks{This work was supported by NSF Award ECCS-2013779.}
}

\newcommand{\n}{{\rm n}}
\renewcommand{\v}{{\rm v}} 
\newcommand{\vu}{{\rm vu}}
\newcommand{\vs}{{\rm vs}}

\newcommand{\type}{\mathcal{T}}
\newcommand{\bayesianType}{\type_B}
\newcommand{\nonBayesianType}{\type_I}

\newcommand{\C}{{\rm C}}
\newcommand{\T}{{\rm T}}
\newcommand{\R}{{\rm R}}

\newcommand{\St}{S_\tau}
\newcommand{\Sn}{S_\n}
\newcommand{\Sv}{S_\v}
\newcommand{\Svu}{S_\vu}
\newcommand{\Svs}{S_\vs}

\newcommand{\Pn}{P_{\rm n}}
\newcommand{\Pvu}{P_{\rm vu}}
\newcommand{\Pvuo}{P_{\rm vu1}}
\newcommand{\Pvut}{P_{\rm vu2}}
\newcommand{\Pvs}{P_{\rm vs}}

\newcommand{\xt}{x_\tau}
\newcommand{\xta}{x_\tau^s}
\newcommand{\xn}{x_\n}
\newcommand{\xna}{x_\n^s}
\newcommand{\xv}{x_\v}
\newcommand{\xva}{x_\v^s}
\newcommand{\xvu}{x_\vu}
\newcommand{\xvua}{x_\vu^s}
\newcommand{\xvuat}{x_{\vu 2}^s}
\newcommand{\xvs}{x_\vs}

\newcommand{\Jt}{J_\tau}
\newcommand{\Jn}{J_{\rm n}}
\newcommand{\Jv}{J_{\rm v}}
\newcommand{\Jvu}{J_{\rm vu}}
\newcommand{\Jvs}{J_{\rm vs}}

\newcommand{\A}{{\rm A}}
\renewcommand{\S}{{\rm S}} 
\newcommand{\B}{{\rm B}}

\newcommand{\Pt}{P_\type}
\newcommand{\Ptb}{P_{\bayesianType}}
\newcommand{\Pti}{P_{\nonBayesianType}}
\newcommand{\Pb}{\bar{P}}

\newcommand{\SC}{\mathcal{J}}
\newcommand{\SCt}{\mathcal{J}_\type}
\newcommand{\SCtb}{\mathcal{J}_{\bayesianType}}
\newcommand{\SCti}{\mathcal{J}_{\nonBayesianType}}

\begin{document}

\maketitle
\thispagestyle{empty}
\pagestyle{empty}

\begin{abstract}
Vehicle-to-Vehicle (V2V) communication is intended to improve road safety through distributed information sharing; however, this type of system faces a design challenge: it is difficult to predict and optimize how human agents will respond to the introduction of this information. 
\emph{Bayesian games} are a standard approach for modeling such scenarios; in a Bayesian game, agents probabilistically adopt various types on the basis of a fixed, known distribution.
Agents in such models ostensibly perform Bayesian inference, which may not be a reasonable cognitive demand for most humans.
To complicate matters, the information provided to agents is often implicitly dependent on agent behavior, meaning that the distribution of agent types is a function of the behavior of agents (i.e., the type distribution is \emph{endogenous}).
In this paper, we study an existing model of V2V communication, but relax it along two dimensions: first, we pose a behavior model which does not require human agents to perform Bayesian inference; second, we pose an equilibrium model which avoids the challenging endogenous recursion.
Surprisingly, we show that the simplified non-Bayesian behavior model yields the exact same equilibrium behavior as the original Bayesian model, which may lend credibility to Bayesian models.
However, we also show that the original endogenous equilibrium model is strictly necessary to obtain certain informational paradoxes; these paradoxes do not appear in the simpler exogenous model.
This suggests that standard Bayesian game models with fixed type distributions are not sufficient to express certain important phenomena.

\end{abstract}

\section{Introduction}
\label{sec:intro}
As technology becomes more omnipresent in today's society, technological solutions are being developed for a broad range of applications.
These applications increasingly include areas that have complex interactions with human society, such as the Internet of Things (IoT) and smart infrastructure concepts like vehicle-to-vehicle (V2V) communication. 

These interactions present a unique challenge to engineers, as prior work has shown that naively implemented solutions, even those that intuitively seem helpful, can unintentionally worsen the problems they were designed to solve~\cite{gould_information_2023, Brown2020a}.
In particular, we consider the context of a traffic congestion game, where it is commonly known that selfish individual behavior is not socially optimal~\cite{ dafermos_traffic_1984, wardrop_road_1952, massicot_public_2019,gairing_selfish_2008, wu_information_2019}. 

Prior work has considered various mechanisms to influence agents to choose socially optimal behaviors, such as financially incentivizing desired behaviors~\cite{Ferguson2021, lazar_learning_2021}. 
\emph{Bayesian persuasion} attempts to influence agents through information design, by strategically revealing or concealing information to change the posterior beliefs of these agents~\cite{kamenica_bayesian_2011, bergemann_information_2019, akyol_information-theoretic_2017, shah_optimal_2022}.
For example, one goal of V2V technology is to improve driver safety by broadcasting warning signals when a road hazard is encountered. 
When a driver receives such a warning (even if it is known to occasionally be incorrect), they have a higher degree of belief that the road is unsafe, and are therefore more likely to drive carefully. 
However, there are significant limitations to information design. 


First, Bayesian persuasion places stringent rationality assumptions on human agents. 
Human decision making is not consistently utility-maximizing~\cite{basu_travelers_1994, tversky_judgment_1974, aumann_rationality_1997}, and can be affected by personal biases~\cite{hebbar_role_2022, gigerenzer_homo_2009} or even by different representations of equivalent information~\cite{kahneman_prospect_1979, edwards_prospect_1996}.
Furthermore, formal statistical statements, and in particular Bayes' Theorem, are often misunderstood by humans, even those in highly educated positions such as doctors~\cite{gigerenzer_helping_2007}. 
Assuming that human agents can quickly and accurately perform this calculation is likely unrealistic. 

Second, it is often non-trivial to design the information sharing policy.  
Prior work has shown that full information sharing is not always optimal, and may even be worse than no information sharing ~\cite{liu_effects_2016, massicot_public_2019, sayin_hierarchical_2019, tavafoghi_informational_2017,bergemann_information_2019}.
A compounding factor in conducting this analysis is model complexity: in many applications, the information to be shared is implicitly a function of agent behavior; this adds additional complexity to the optimization problem.

Our work is designed to address both of these issues, using the context of a congestion game where V2V cars are able to share information about accidents, previously introduced in~\cite{gould_information_2023}. 
We begin by posing a novel model of agent decision-making which does not require agents to perform Bayesian inference.
Surprisingly, we show that in this non-Bayesian model, equilibrium behavior exactly matches that of the original Bayesian model.
This transformation allows us to relax the rationality expectations on human drivers, potentially improving the real-world descriptive power of the model. 
In addition, it allows us to re-frame the original model of a Bayesian game of incomplete information to a non-Bayesian game of imperfect information, potentially opening new avenues for analysis.
This is reminiscent of Harsanyi's classic work~\cite{harsanyi_games_1967}, but we believe that our characterization is not a direct consequence of this prior work.
In particular, our setting has non-atomic agents, and we allow the distribution of agent types to vary endogenously, neither of which is considered in~\cite{harsanyi_games_1967}. 

Next, we investigate the relationship between model complexity and expressiveness by considering two classes of models.
The simpler approach assumes that the probability of an accident (and thus the distribution of which agents receive which types of signals) is a constant model parameter, unaffected by an emergent behavior (i.e. it is exogenous). 
Note that this is the standard approach in information design problems in the literature; road hazards or highway delay characteristics are almost universally assumed to be drawn from some fixed, known distribution~\cite{gairing_selfish_2008,massicot_public_2019,wu_information_2019,Zhu2020,Ferguson2022}.

However, one would intuitively expect the probability of an accident to depend on the behavior of drivers, where more reckless behaviors make accidents more likely. 
Therefore,~\cite{gould_information_2023} considers models with an endogenous accident probability, expressing it as a function of equilibrium behavior. 
This creates a complex recursive relationship where accident probability varies in response to driver behavior, which varies in response to accident probability. 

In the present paper, we show that even though the simpler exogenous models are easier to analyze, they are qualitatively different and cannot describe the same phenomena as the endogenous models. 
In particular, when the accident probability is endogenous, a paradox can occur in which social cost increases with information sharing.
On the other hand, this paradox never occurs in the simpler (more standard) exogenous model. 
This suggests that in some circumstances, the current popular modeling framework of Bayesian games with fixed agent type distributions may be insufficient to express important phenomena.

\section{Model}
\label{sec:model} 

\subsection{General Setup}
\label{subsec:generalSetup}
Our model consists of a non-atomic, unit mass of agents (drivers) interacting on a single road. 
On this road, traffic accidents either occur ($\A$) or do not occur ($\neg \A$). 
Throughout, we use $\mathbb{P}(E)$ to represent the probability of event $E$. 

Drivers are able to choose between the actions of driving carefully ($\C$) or recklessly ($\R$).
Intuitively, careful drivers consistently choose slower, safer driving behaviors such as signaling before changing lanes, while reckless drivers choose faster, riskier behaviors. 
Reckless drivers become involved in existing accidents and experience an expected cost of $r>1$; however, careful drivers regret their caution if an accident is not present and experience a regret cost of $1$.
These costs are collected in this matrix:

\begin{center}
	\begin{game}{2}{2}
		\relax&Accident ($\A$)&No Accident ($\neg \A$)\\
		Careful ($\C$)&$0$&$1$\\
		Reckless ($\R$)&$r$&$0$\\
	\end{game}
\end{center}

Each driver has type $\tau \in \type$, and a set of strategies $\St$.
A \emph{strategy} $s \in \St$ for a driver is a procedure to choose which action to play as a function of the information known to the agent. 
Let $\xta$ represent the mass of drivers of type $\tau$ choosing strategy $s \in \St$, and all agents' behavior is described by the tuple $x = (\xta)_{\tau \in \type, s \in \St}$.
We use $\xt$ for the total mass of drivers of type $\tau$, so that $\sum_{s \in \St} \xta = \xt$. 

A fraction $y$ of agents drive cars equipped with V2V communication technology. 
Cars with this technology can autonomously detect accidents, and broadcast signals about them. 
If an accident occurs, a ``true positive'' signal is broadcast with probability $t(y)$; otherwise, a ``false positive'' signal is broadcast with probability $f(y) < t(y)$. 
Any signals that are broadcast are received by all V2V drivers. 

Counter-intuitively, sharing perfect information about the environment can make parts or all of the population worse off. 
Therefore, it may sometimes be optimal for administrators of V2V technology to withhold information from drivers; \cite{gould_information_2023} showed that this is the case for a specific instantiation of a much more general class of models considered in this work. 

Accordingly, we introduce the parameter $\beta$ to describe the information quality of V2V technology. 
In the event that a warning signal is broadcast $(\B)$, a V2V car may not always display a warning signal $(\S)$ to its driver; it will do so with probability $\beta = \mathbb{P}(\S | \B) \in [0, 1]$.
Therefore, we have that 
\begin{equation}
    \label{eq:signalProbDef}
    \mathbb{P}(\S) = \beta(\mathbb{P}(\A)t(y) + (1-\mathbb{P}(\A))f(y)).
\end{equation}
The system planner performs \emph{information design} on the value of $\beta$ to minimize accident probability and social cost. 

An attractively simple approach to analysis is to let the probability of an accident be a constant that is unaffected by social behavior; this approach has been used previously in the literature~\cite{lazar_learning_2021, wu_information_2019}. 
We call this case an \emph{exogenous} accident probability, and refer to it as simply $ \mathbb{P}(\A) = \Pb \in [0, 1]$. 

However, this assumption seems inaccurate to the real world, since we expect that reckless driving habits would lead to more frequent accidents. 
Therefore, we wish to allow accident probability to be endogenously affected by driver behavior. 
To that end, we write $d \in [0, 1]$ to denote the overall fraction of drivers choosing to drive recklessly, and $p(d)$ for the resulting probability that an accident occurs. 
We assume that $p$ is a continuous, strictly increasing function, so that more reckless drivers cause accidents to be more likely. 

Intuitively, strategies in which drivers choose to drive recklessly more often should cause accidents to become more likely; to measure this, let $\rho(\tau, s, P)$ be the probability that a driver of type $\tau \in \type$ choosing strategy $s \in \St$ is reckless when the probability of an accident is $P$. 
Then, the mass of reckless drivers of type $\tau$ choosing strategy $s$ when accident probability is $P$ is $\rho(\tau, s, P) \xta$.

Then, the endogenous accident probability resulting from a behavior tuple $x$ for driver types $\type$ is described implicitly as a solution to the recursive relationship:
\begin{equation}
    \label{eq:generalPRecursion}
    \Pt(x) = p\left(\sum_{\tau \in \type} \sum_{s \in \St} \rho(\tau, s, \Pt(x)) \xta \right).
\end{equation}
It was shown in~\cite[Proposition 2.1]{gould_information_2023} this this relationship always has such a solution. 

We consider two classes of games, distinguished by either exogenous or endogenous accident probabilities. 
We write the former as the tuple $\bar{G} =(\beta, y, r, \Pb)$, and the latter as $G = (\beta, y, r)$. 
For both types of games, equilibrium conditions come from the standard Nash idea: a behavior tuple $x$ is an equilibrium of $G$ if for any type $\tau$ and any strategy $s \in \St$, 
\begin{equation}
    \label{eq:ActionInUseCond}
    \xta > 0 \implies \Jt(s; x) = \min\limits_{s \in \St} \Jt(s; x).
\end{equation}
That is, if any driver is choosing a strategy, its cost to them is minimal. 

Finally, we define social cost as the expected cost incurred by the entire population: 
\begin{equation}
    \label{eq:socialCostDef}
    \SCt(x) = \sum_{\tau\in \type}\sum_{s \in \St} \Jt(s, x)\xta.
\end{equation}
We abuse notation and write $\Pt(G)$ to mean $\Pt(x)$ and $\SCt(G)$ (or $\SCt(\bar{G})$) to mean $\SCt(x)$ where $x$ is an equilibrium of the game $G$ (or $\bar{G}$).%
\footnote{It can be shown that any two equilibria of the same game have equal accident probability and social cost.
An intuitive explanation for this is: every game has a unique equilibrium unless at least one type of drivers is indifferent between multiple strategies. 
If this is the case, then these strategies have equal cost, meaning the relative frequencies of drivers choosing each strategy are in some sense irrelevant. 
See the proof of Lemma~\ref{lemma:rationalExoSocialCost} for a formal treatment of this idea in exogenous games, and Lemma~\ref{lemma:subrationalEndoEqUniqueness} for the same in endogenous games. 
}

\subsection{Driver Decision Models}
\label{subsec:decisionTypes}
We consider two interpretations for the effect of V2V technology on the behavior of V2V drivers. 

\subsubsection{Bayesian Agents} The first is a more standard approach for a Bayesian game, and is studied in Section~\ref{sec:rationalAgents} (additionally, \cite{gould_information_2023} analyzes a specific case of this approach in great detail).
There are three types of agents: non-V2V drivers, V2V drivers who receive a signal, and V2V drivers who do not receive a signal. 
Non-V2V drivers must use the prior probability of an accident to calculate their expected costs, but we allow V2V drivers to use the posterior probability after the signal realization. 

We call this the \emph{Bayesian} set of types, and write it $\bayesianType = \{\n, \vu, \vs\}$ for non-V2V, unsignaled V2V, and signaled V2V drivers respectively.
It can be quickly seen that the mass of drivers in each group is $\xn = 1-y$, $\xvu = \mathbb{P}(\neg \S)y$, and $\xvs = \mathbb{P}(\S)y$.
Each type has the strategies $\Sn = \Svu = \Svs = \{\C, \R\}$, with the accompanying cost functions: 
\begin{align}
    \Jn (s;x)&=\begin{cases}
        1-\mathbb{P}(\A) & \text{if $s=\C$},\\
        r\mathbb{P}(\A) & \text{if $s=\R$},
        \label{eq:RationalNV2VCost}
    \end{cases} \\
    \Jvu (s;x)&=\begin{cases}
        1 - \mathbb{P}(\A | \neg \S) & \text{if $s=\C$},\\
        r \mathbb{P}(\A | \neg \S) & \text{if $s=\R$},
        \label{eq:RationalV2VUnsignaledCost}
    \end{cases} \\
    \Jvs (s;x)&=\begin{cases}
        1 - \mathbb{P}(\A | \S) & \text{if $s=\C$},\\
        r \mathbb{P}(\A | \S) & \text{if $s=\R$}.
        \label{eq:RationalV2VSignaledCost}
    \end{cases}
\end{align}

\subsubsection{Non-Bayesian Agents} Alternatively, we consider a decision model where V2V drivers do not perform a Bayesian update (studied in Section~\ref{sec:subrationalAgents}). 
In this case, the only driver types are non-V2V and V2V, written $\nonBayesianType = \{\n, \v\}$ with $\xn = 1-y$ and $\xv = y$.
Instead of calculating a posterior, V2V drivers choose between trusting ($\T$) the signal completely (assuming that the presence of a signal implies the existence of an accident, and vice versa), or ignoring it completely, using the prior probability to calculate an ``assumed'' cost of driving carefully ($\C$) or recklessly ($\R$). 

Drivers have strategies $\Sn = \{\C, \R\}$ and $\Sv = \{\T, \C, \R\}$. 
The cost functions for these strategies are: 
\begin{align}    
    \Jn(s; x) &= \begin{cases}
        1 - \mathbb{P}(\A)  &\text{ if $s=\C$}, \\
        r\mathbb{P}(\A)  &\text{ if $s=\R$}.
        \label{eq:NV2VSubrationalAverageCost}
    \end{cases} \\
    \Jv(s; x) &= \begin{cases}
        \mathbb{P}(\neg \A \cap \S) + r\mathbb{P}(\A \cap \neg \S) & \text{ if } s = \T, \\
        1 - \mathbb{P}(\A)  &\text{ if $s=\C$}, \\
        r\mathbb{P}(\A)  &\text{ if $s=\R$}.
        \label{eq:V2VSubrationalAverageCost}
    \end{cases}
\end{align}

\begin{figure*}
    \centering
    \includegraphics[width=\textwidth]{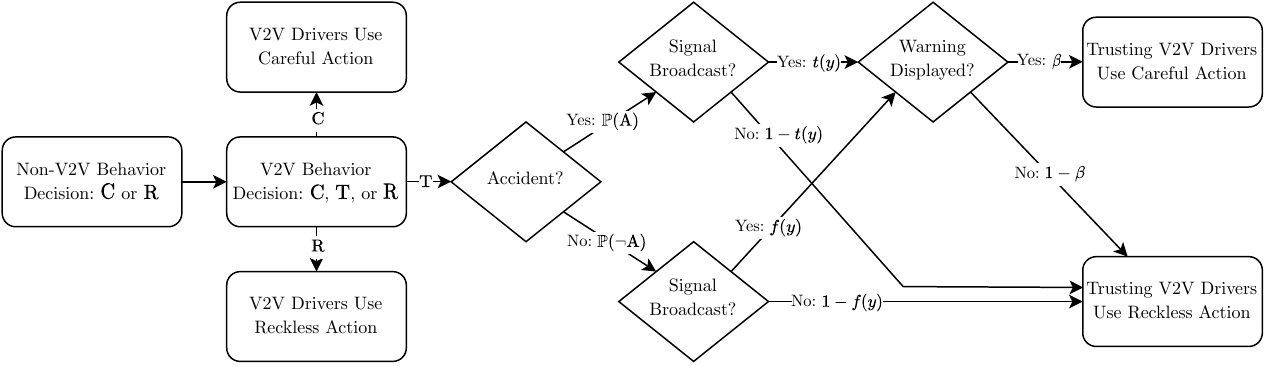}
    \caption{Timeline of decisions and events for non-Bayesian drivers.}
    \label{fig:timeline}
\end{figure*}

The timeline of games with non-Bayesian agents is shown in Figure~\ref{fig:timeline}.
Note the differences between this timeline and that of in~\cite[Figure 1]{gould_information_2023} (describing the behavior of Bayesian agents).
Non-Bayesian V2V agents must commit to a strategy \emph{before} the signal realization, and can only use the information provided by the signal if they commit to trusting it completely. 

\section{Main Results}
Our first result is that Bayesian games of \emph{incomplete} information can be reinterpreted as \emph{imperfect} information games, similar to the transformation in~\cite{harsanyi_games_1967}. 
The benefit of doing so is that it allows us to remove the heavy cognitive burden of Bayes' Theorem from agent calculations, producing a more credible model of human behavior.

\begin{theorem}
    \label{thm:informationTrade}
    Let $G = (\beta, y, r)$ be a game with endogenous accident probability, and $\bar{G} = (\beta, y, r, \Pb)$ be an exogenous game. 
    Then,
    \begin{equation}
        \label{eq:exoSCEquality}
        \SCtb(\bar{G}) = \SCti(\bar{G}),
    \end{equation}
    \centering{and}
    \begin{align}
        \Ptb(G) &= \Pti(G), \label{eq:endoPEquality} \\ 
        \SCtb(G) &= \SCti(G). \label{eq:endoSCEquality}
    \end{align}
\end{theorem}

\begin{proof}
    Equation \eqref{eq:exoSCEquality}, concerning the social cost of games with exogenous accident probability, is proven by Lemma~\ref{lemma:exoSCEquality} in section~\ref{subsec:exoSubrational}. 
    Equations \eqref{eq:endoPEquality} and \eqref{eq:endoSCEquality}, analogous statements for games with endogenous accident probability, are proven by Lemma~\ref{lemma:endoCPEquality} and Lemma~\ref{lemma:endoSCEquality} respectively in section~\ref{subsec:endoSubrational}.
    This completes the proof.
\end{proof}

Intuitively, Theorem~\ref{thm:informationTrade} shows that allowing drivers to use Bayesian inference in their decisions does \emph{not} gain any significant model expressiveness. 
The same outcomes and paradoxes can be captured by a decision model where drivers choose between complete trust and complete ignorance of the signal's information. 
This gives us freedom to analyze any game assuming either Bayesian or non-Bayesian drivers, whichever is more convenient. 
Additionally, since the social costs induced by both models are equivalent, we now write simply $\SC(G)$ for the social cost of the game $G$, where $G$ can be interpreted as either a Bayesian or non-Bayesian game. 

Our second result concerns the modeling of accidents. 
It seems reasonable to expect (and is indeed true) that using an endogenous accident probability significantly complicates model analysis. 
Because of this, a natural question is to ask whether the same characteristics can be captured by a model using exogenous accident probability. 
However, our next result shows that this is not the case. 
It was shown in~\cite{gould_information_2023} that for certain games with endogenous accident probability, equilibrium social cost can paradoxically be increasing with information quality. 
By contrast, no such social cost paradox is possible when using an exogenous accident probability.

\begin{theorem}
    \label{thm:endoNecessity}
    Fix $y\in [0, 1]$, $r > 1$, and $\Pb \in [p(0), p(1)]$. 
    Consider $\beta_1, \beta_2 \in [0, 1]$ with $\beta_1 < \beta_2$.
    Let $\bar{G}_1 = (\beta_1, y, r, \Pb)$ and $\bar{G}_2 = (\beta_2, y, r, \Pb)$ be signaling games with exogenous accident probability. 
    Then, it is always true that 
    \begin{equation}
        \label{eq:exoSCNondecreasing}
        \SC(\bar{G}_1) \ge \SC(\bar{G}_2).
    \end{equation}
    
    Similarly, Let $G_1 = (\beta_1, y, r)$ and $G_2 = (\beta_2, y, r)$ be signaling games with endogenous accident probability.
    Counter-intuitively, there exist parameter combinations where
    \begin{equation}
        \label{eq:endoSCDecreasing}
        \SC(G_1) < \SC(G_2). 
    \end{equation}
\end{theorem}

\begin{proof}
    We first consider games between Bayesian drivers.
    In this case, Lemma~\ref{lemma:rationalExoSocialCost} proves \eqref{eq:exoSCNondecreasing}, and Lemma~\ref{lemma:rationalEndoSocialCost} shows that \eqref{eq:endoSCDecreasing} can be satisfied. 

    Then, by Theorem~\ref{thm:informationTrade}, these results also hold for games between non-Bayesian drivers. 
    This completes the proof. 
\end{proof}

\section{Bayesian Agents}
\label{sec:rationalAgents} 
We first consider a scenario where V2V drivers perform Bayesian updates on the probability of an accident using the warning signal realization.
Analysis proceeds as follows: Lemma~\ref{lemma:rationalStrictBehavior} establishes necessary conditions of any equilibrium (with either exogenous or endogenous accident probability) by describing behavior when drivers have a strict preference for one strategy. 
This result is sufficient to show that social cost is monotonically decreasing with information quality for games with exogenous accident probability, which we do in section~\ref{subsec:exoRational} through Lemma~\ref{lemma:rationalExoSocialCost}. 
Finally, in section~\ref{subsec:endoRational}, Lemma~\ref{lemma:rationalEndoSocialCost} gives the counter-intuitive result that social cost may sometimes be increasing with information quality, but only if accident probability is endogenous. 

\subsection{Necessary Equilibrium Conditions}
\label{subsec:rationalEQNecessary}
We divide V2V drivers into two groups by whether or not they have seen a warning signal. 
This implies the set of driver types $\bayesianType = \{\n, \vu, \vs\}$.
Using the calculated posterior probability, V2V drivers choose between the pure strategies of always driving carefully or always driving recklessly. 

The following thresholds are useful in our analysis; we define the following shorthand to reference them: 
\begin{align}
    \Pvs &:= \frac{f(y)}{rt(y)+f(y)},  \label{eq:V2VsignaledThreshold} \\ 
    \Pn &:= \frac{1}{1+r}, \\ 
    \Pvu &:= \frac{1-\beta f(y)}{1 + r(1-\beta t(y)) -\beta f(y)}, \label{eq:V2VunsignaledThreshold} 
\end{align}
where it holds that:
\begin{equation}
    \label{eq:recklesnessOrdering}
    \Pvs < \Pn \leq \Pvu.
\end{equation}

Note that (by Bayes' Theorem and \eqref{eq:RationalNV2VCost}-\eqref{eq:RationalV2VSignaledCost}):
\begin{align}
    \begin{split}
        \label{eq:V2VUnsignaledBayes}
        \Jvu(\R; x) 
        &\def\arraystretch{0.5}\begin{array}{c}<  \\ = \\ >\end{array}
        \Jvu(\C; x) \iff 
        \mathbb{P}(\A)
        \def\arraystretch{0.5}\begin{array}{c}< \\ = \\ >\end{array}
        \Pvu,
    \end{split} \\
    \begin{split}
        \label{eq:V2VSignaledBayes}
        \Jvs(\R; x) 
        &\def\arraystretch{0.5}\begin{array}{c}<  \\ = \\ >\end{array}
        \Jvs(\C; x) \iff 
        \mathbb{P}(\A)
        \def\arraystretch{0.5}\begin{array}{c}< \\ = \\ >\end{array}
        \Pvs
    \end{split}
     \def\arraystretch{1.0}
\end{align}

We use this notation to mean that any of the relationships between the first expressions is equivalent to the corresponding relationship between the later expressions. 
Equality and both inequalities are preserved. 

Using \eqref{eq:RationalNV2VCost}-\eqref{eq:RationalV2VSignaledCost}, we are now equipped to state the necessary conditions of any equilibrium: 
\begin{lemma}
\label{lemma:rationalStrictBehavior} For any equilibrium $x$ of a game with either endogenous or exogenous accident probability, 
\begin{align}
    \xn^\R &= \begin{cases}
        0 & \text{if } \mathbb{P} (\A) > \Pn, \\
        1-y & \text{if } \mathbb{P} (\A) < \Pn,
        \label{eq:NV2VChoice}
    \end{cases} \\
    \xvu^\R &= \begin{cases}
        0 & \text{if }\mathbb{P} (\A) > \Pvu, \\
        \mathbb{P}(\neg \S)y & \text{if } \mathbb{P} (\A) < \Pvu,
        \label{eq:V2VUnsignaledChoice}
    \end{cases} \\
    \xvs^\R &= \begin{cases}
        0 & \text{if } \mathbb{P} (\A) > \Pvs, \\
        \mathbb{P}(\S)y & \text{if } \mathbb{P} (\A) < \Pvs.
        \label{eq:V2VSignaledChoice}
    \end{cases}
\end{align}
\end{lemma}

\begin{proof}
Each of these implications follows immediately from \eqref{eq:ActionInUseCond}, \eqref{eq:RationalNV2VCost}-\eqref{eq:RationalV2VSignaledCost}, and \eqref{eq:V2VUnsignaledBayes}-\eqref{eq:V2VSignaledBayes}. 

Assume by way of contradiction that $\mathbb{P}(\A) > \Pvu$, but $\xvu^\R > 0$. 
By \eqref{eq:V2VUnsignaledBayes}, $\mathbb{P}(\A | \neg \S) > \frac{1}{1+r}$. 
Then, \eqref{eq:RationalV2VUnsignaledCost} gives that 
\[\
    Jvu(\R, x) = r \mathbb{P}(\A | \neg \S) > 1 - \mathbb{P}(\A | \neg \S) = \Jvu(\C, x).
\]
But this clearly contradicts \eqref{eq:ActionInUseCond}, since $\xvu^\R > 0$.

A proof of the remaining cases is accomplished in a similar manner. 
\end{proof}

\subsection{Exogenous Crash Probability}
\label{subsec:exoRational}
We first assume that accident probability is defined as an exogenous constant $\Pb$.
Since accident probability is constant, it is very straightforward to compute social cost at equilibrium using \eqref{eq:socialCostDef}. 
We claim that this social cost is non-increasing with information quality $\beta$.

\begin{lemma}
\label{lemma:rationalExoSocialCost}
Let $\beta_1, \beta_2 \in [0, 1]$ with $\beta_1 < \beta_2$.
Consider games $\bar{G}_1 = (\beta_1, y, r, \Pb)$ and $\bar{G}_2 = (\beta_2, y, r, \Pb)$ between Bayesian agents with exogenous accident probability $\Pb$. 
Then, $\SCtb(\bar{G}_1) \ge \SCtb(\bar{G}_2)$. 
\end{lemma}

\begin{proof}
    Let $\Pvuo$ and $\Pvut$ represent the threshold $\Pvu$ for $\beta_1$ and $\beta_2$, respectively, and note that $\Pvuo < \Pvut$ since $f(y) < t(y)$. 
    
    First, assume that $\Pb > \Pvut > \Pvuo$.
    By \eqref{eq:recklesnessOrdering}, $\Pb > \Pn > \Pvs$. 
    By Lemma~\ref{lemma:rationalStrictBehavior}, we then have that $\xn^\R = 0$, $\xvu^\R = 0$, and $\xvs^\R = 0$ for any equilibrium of $\bar{G}_1$ or $\bar{G}_2$. 

    Thus, \eqref{eq:RationalNV2VCost}-\eqref{eq:RationalV2VSignaledCost} substitute into \eqref{eq:socialCostDef} to give
    \[
        \SCtb(\bar{G}_1) = 1-\Pb \ge \SCtb(\bar{G}_2), 
    \]
    and we are finished in this case. 

    Now, assume that $\Pb = \Pvut$ (implying $\Pb > \Pvuo$). 
    By \eqref{eq:recklesnessOrdering}, $\Pvs < \Pn < \Pb$, so by Lemma~\ref{lemma:rationalStrictBehavior}, $\xn^\R = 0$ and $\xvs^\R = 0$ in any equilibrium of $\bar{G}_1$ or $\bar{G}_2$. 
    Let $x_1$ be an equilibrium of $\bar{G}_1$ and $x_2$ one of $\bar{G}_2$. 
    By the above, $\SCtb(\bar{G}_1) = 1-\Pb$. 
    
    Since $\Pb = \Pvut$, \eqref{eq:RationalV2VSignaledCost} and simple algebra give that $\Jvu(\C, x_2) = \Jvu(\R, x_2)$. 
    Thus, $\sum_{s \in \Svu} \Jvu(s, x_2)\xvuat = \Jvu(\R, x)\mathbb{P}(\neg \S)y$. 
    Then, \eqref{eq:socialCostDef} simplifies to 
    \[
        \SCtb(\bar{G}_1) = 1 - \Pb \ge 1 - \Pvut = \SCtb(\bar{G}_2),
    \]
    and we are again finished. 
    A similar technique is used when $\Pvuo = \Pb < \Pvut$, $\Pn < \Pb < \Pvuo$, $\Pb = \Pn$, $\Pvs < \Pb < \Pn$, $\Pb = \Pvs$, or $\Pb < \Pvs$, completing the proof. 
\end{proof}

Lemma~\ref{lemma:rationalExoSocialCost} shows that providing information of a higher quality will never increase social cost when accident probability is exogenous, which is perhaps the expected result. 
However, this result no longer holds in a model with an endogenous accident probability. 

\subsection{Endogenous Crash Probability}
\label{subsec:endoRational}
We now consider a model of Bayesian drivers with an endogenous accident probability. 
Recall that agents have one of three types: non-V2V drivers, unsignaled V2V drivers, and signaled V2V drivers. 
Drivers of each type $\tau \in \bayesianType$ have access only to the pure strategies of always driving carefully or always driving recklessly, where $\rho(\tau, \R, \Ptb(x)) = 1$ and $\rho(\tau, \C, \Ptb(x)) = 0$.

Therefore,~\eqref{eq:generalPRecursion} simplifies to 
\begin{equation}
    \label{eq:rationalPRecursion}
    \mathbb{P}(\A) = \Ptb(x) = p\left(\xn^\R + \xvu^\R + \xvs^\R\right).
\end{equation}

In this case, the model is identical to the one described in \cite{gould_information_2023}. 
Since it was thoroughly studied in that work, we simply reference the previous result. 

\begin{lemma}[\hspace{-.1mm}{\cite[Proposition 3.8]{gould_information_2023}}]
    \label{lemma:rationalEndoSocialCost}
    There exist games between Bayesian drivers with endogenous accident probability such that social cost at equilibrium is increasing with $\beta$. 
\end{lemma}

\section{Non-Bayesian Agents}
\label{sec:subrationalAgents}
We now discuss games with agents who do not perform explicit Bayesian updates. 
Analysis generally follows the same path as section~\ref{sec:rationalAgents}, with some additional components. 
Lemma~\ref{lemma:subrationalStrictBehavior} describes the behavior of agents who have a strict preference for one strategy, giving necessary conditions for an equilibrium of any game between non-Bayesian agents, whether accident probability is exogenous or endogenous. 

We then focus on games with exogenous accident probability. 
Lemma~\ref{lemma:subrationalExoSCNonIncreasing} shows that for such games, social cost is decreasing with information quality. 
Then, Lemma~\ref{lemma:exoSCEquality} compares the consequences of Bayesian and non-Bayesian agents in these games, and shows that equilibrium social cost is the same regardless of which agent behaviors are used. 

Finally, we consider games with endogenous accident probability. 
Lemmas~\ref{lemma:subrationalEndoEqUniqueness} and \ref{lemma:subrationalEndoParameterCrashProbMapping} characterize equilibria of these games; the former describes how driver behavior is determined by accident probability, and the latter gives a relationship between game parameters and accident probability. 
With this, driver behavior and equilibrium accident probability can be determined exclusively from game parameters. 
Finally, Lemmas~\ref{lemma:endoCPEquality} and \ref{lemma:endoSCEquality} compare the accident probability and social cost induced by Bayesian agents to those of non-Bayesian agents, and find that there is no difference. \looseness=-1 

\subsection{Necessary Equilibrium Conditions}
\label{subsec:subrationalEQNecessary}
To model the behavior of non-Bayesian agents, we use the alternate set of driver types $\nonBayesianType = \{\n, \v\}$, representing non-V2V drivers and V2V drivers, respectively. 
V2V agents are still able to condition their behavior on the realization of a warning light, but must either completely trust or ignore the signal, and use an ``assumed'' cost equal to that of non-V2V drivers for decision making when ignoring the signal. 

Again, we use thresholds to describe behavior. 
By \eqref{eq:V2VSubrationalAverageCost}, 
\begin{align}
    \label{eq:V2VTrustCareful}
    \Jv(\T; x) 
    \def\arraystretch{0.5}\begin{array}{c}<  \\ = \\ >\end{array}
    \Jv(\C; x) &\iff 
    \mathbb{P}(\A)
    \def\arraystretch{0.5}\begin{array}{c}< \\ = \\ >\end{array}
    \Pvu, \\
    \label{eq:V2VTrustReckless}
    \Jv(\R; x) 
    \def\arraystretch{0.5}\begin{array}{c}<  \\ = \\ >\end{array}
    \Jv(\T; x) &\iff
    \mathbb{P}(\A)
    \def\arraystretch{0.5}\begin{array}{c}< \\ = \\ >\end{array}
    \Pvs
     \def\arraystretch{1.0}
\end{align}

Crucially, Bayes' Theorem was never necessary for these calculations, yet they give the \emph{same} behavior thresholds as \eqref{eq:V2VUnsignaledBayes} and \eqref{eq:V2VSignaledBayes} in  Section~\ref{sec:rationalAgents}. 
This gives a very compelling argument for our claim that Bayesian and non-Bayesian behaviors are equivalent. 
Once it is known that both models are in some sense making the same decisions, it is unsurprising that the resulting equilibria are identical. 
We now make an analogous statement to Lemma~\ref{lemma:rationalStrictBehavior} using \eqref{eq:V2VTrustCareful} and \eqref{eq:V2VTrustReckless}. 

\begin{lemma}
    \label{lemma:subrationalStrictBehavior} 
    For any equilibrium of $x$ of a game, 
    \begin{align}
    \xn^\R &= \begin{cases}
        0 & \text{if } \mathbb{P} (\A) > \Pn, \\
        1-y & \text{if } \mathbb{P} (\A) < \Pn,
        \label{eq:NV2VChoice2}
        \end{cases} \\
    \xv^\R &= \begin{cases}
        0 & \text{if } \mathbb{P} (\A) > \Pvs, \\
        y & \text{if } \mathbb{P} (\A) < \Pvs,
        \label{eq:V2VRecklessChoice}
        \end{cases} \\
    \xv^\T &= \begin{cases}
        0 & \text{if } \mathbb{P} (\A) < \Pvs, \\
        y & \text{if } \Pvs < \mathbb{P} (\A) < \Pvu, \\
        0 & \text{if } \mathbb{P} (\A) > \Pvu
        \label{eq:V2VTrustingChoice}
        \end{cases}
    \end{align}
\end{lemma}

\begin{proof}
    This follows directly from \eqref{eq:ActionInUseCond}, \eqref{eq:NV2VSubrationalAverageCost}-\eqref{eq:V2VSubrationalAverageCost}, and \eqref{eq:V2VTrustCareful}-\eqref{eq:V2VTrustReckless}.
    
    Assume by way of contradiction that $\Pvs < \mathbb{P}(\A) < \Pvu$, but $\xv^\T < y$. 
    By \eqref{eq:V2VTrustCareful}, $\Jv(\T; x) < \Jv(\C; x)$, and by \eqref{eq:V2VTrustReckless}, $\Jv(\T; x) < \Jv(\R; x)$. 
    Therefore, by \eqref{eq:ActionInUseCond}, $\xv^\C = \xv^\R = 0$. 
    But since $\xv = y$, this implies $\xv^\T = y$, a contradiction. 

    A proof of the remaining cases can be accomplished in a similar manner. 
\end{proof}

\subsection{Exogenous Crash Probability}
\label{subsec:exoSubrational}
As before, we begin analysis by assuming an exogenous, constant accident probability $\Pb$.
Again, we see that models making this assumption are qualitatively different from those allowing for an interdependence between driver behavior and accident probability. 

\begin{lemma}
\label{lemma:subrationalExoSCNonIncreasing} 
Let $\beta_1, \beta_2 \in [0, 1]$ with $\beta_1 < \beta_2$.
Consider games $\bar{G}_1 = (\beta_1, y, r, \Pb)$ and $\bar{G}_2 = (\beta_2, y, r, \Pb)$ between Bayesian agents with exogenous accident probability $\Pb$. 
Then, $\SCti(\bar{G}_1) \ge \SCti(\bar{G}_2)$.
\end{lemma}

\begin{proof}
    This can be shown using a technique identical to that of the proof of Lemma~\ref{lemma:rationalExoSocialCost}.
\end{proof}

That is, social cost is again non-increasing with information quality $\beta$ for these types of games. 
Furthermore, we show in Lemma~\ref{lemma:exoSCEquality} that social cost is unaffected by the change in the decision model of V2V drivers:

\begin{lemma}
\label{lemma:exoSCEquality}
For any game $\bar{G} = (\beta, y, r, \Pb)$ with exogenous accident probability $\Pb$, 
\begin{equation}
    \SCtb(\bar{G}) = \SCti(\bar{G}). 
\end{equation}
\end{lemma}

\begin{proof}
We again prove this in cases. 
First, assume that $\Pn < \Pb < \Pvu$. 
By Lemma~\ref{lemma:rationalStrictBehavior}, $\xn^\R = 0$, $\xvu^\R = \mathbb{P}(\neg \S)y$, and $\xvs^\R = 0$. 
Similarly by Lemma~\ref{lemma:subrationalStrictBehavior}, $\xn^\R = 0$, $\xv^\R = 0$, and $\xv^\T = y$ (note that the value of $\xn^\R$ is unchanged regardless of which of the two Lemmas is used to derive it). 
Then, \eqref{eq:socialCostDef} simplifies to give 
\begin{align*}
    \SCtb&(\bar{G}) = \SCti(\bar{G}) = \\ 
    &(1-\Pb)(1-y) + r\Pb (1-\beta t(y)) y + ((1-\Pb)\beta f(y))y,
\end{align*}

completing the proof in this case. 
The same technique gives the desired result if $\Pb < \Pvs$, $\Pvs < \Pb < \Pn$, or $\Pvu < \Pb$. 

Now, consider the case where $\Pb = \Pn$. 
By Lemma~\ref{lemma:rationalStrictBehavior}, $\xvs^\R = 0$, and by Lemma~\ref{lemma:subrationalStrictBehavior}, $\xv^\R = 0$.
By \eqref{eq:RationalNV2VCost} (or equivalently \eqref{eq:NV2VSubrationalAverageCost}), $\Jn(\R; x) = \Jn(\C; x)$, so $\sum_{s \in \Sn} \Jn(s, x)\xna = \Jn(\R, x)(1-y)$. 

If $\bar{P} = \Pvu$, then a very similar argument implies that $\sum_{s \in \Svu} \Jvu(s, x)\xvua = \Jvu(\R, x)(1-\mathbb{P}(\S))y$, and $\sum_{s \in \Sv} \Jv(s, x)\xva = \Jv(\T, x)y$.
Otherwise, by \eqref{eq:recklesnessOrdering}, $\Pvs < \Pb < \Pvu$, meaning $\xvu^\R = (1-\mathbb{P}(\S))y$ by Lemma~\ref{lemma:rationalStrictBehavior} and $\xv^\T = y$ by Lemma~\ref{lemma:subrationalStrictBehavior}. 
In any case, \eqref{eq:socialCostDef} again simplifies to
\begin{align*}
    \SCtb&(\bar{G}) = \SCti(\bar{G}) = \\ 
    &(1-\Pb)(1-y) + r\Pb (1-\beta t(y)) y + ((1-\Pb)\beta f(y))y,
\end{align*}
which is the desired result. 
The above technique also suffices in the cases where $\Pb = \Pvs$ and $\Pb = \Pvu$, completing the proof. 
\end{proof}

\subsection{Endogenous Crash Probability}
\label{subsec:endoSubrational}
Finally, we introduce an endogenous accident probability to the non-Bayesian decision model. 
Both non-V2V and V2V drivers have access to the pure strategies of always being careful and always being reckless (using the prior probability of an accident to compute expected cost), but V2V drivers additionally have the option to fully ``trust'' the signal, believing the presence of a warning light implies the existence of an accident and the inverse. 

This creates the strategy spaces $\Sn = \{\C, \R\}$ and $\Sv = \{\T, \C, \R\}$, where $\rho(\tau, \R, \Pti(x)) = 1$ and $\rho(\tau, \C, \Pti(x)) = 0$ for each $\tau \in \nonBayesianType$, and $\rho(\xv, \T, \Pti(x)) = \mathbb{P}(\neg \S)$.
Then, \eqref{eq:generalPRecursion} simplifies to 
\begin{equation}
    \label{eq:subrationalPRecursion}
    \mathbb{P}(\A) = \Pti(x) = p(\xn^\R + \xv^\R + \mathbb{P}(\neg\S)\xv^\T).
\end{equation}

Recall that Lemma~\ref{lemma:subrationalStrictBehavior} specifies the equilibrium behavior when agents have a strict preference for one strategy. 
We now provide a description of this behavior when agents are indifferent between two strategies. 

\begin{lemma}
\label{lemma:subrationalEndoEqUniqueness}
For any game $G = (\beta, y, r)$, a non-Bayesian behavior tuple $x$ is a signaling equilibrium if it satisfies Lemma~\ref{lemma:subrationalStrictBehavior} and the following hold: 

\begin{align}
    \mathbb{P}(\A) = \Pn &\implies \xn^\R = p^{-1}(\Pn) - \mathbb{P}(\neg \S)y \label{eq:subrationalNV2VIndif} \\
    \mathbb{P}(\A) = \Pvs &\implies \xv^\R = y \label{eq:subrationalV2VIndifR} \\
    \mathbb{P}(\A) = \Pvs &\implies \xv^\T = 0 \label{eq:subrationalV2VIndifT} \\
    \mathbb{P}(\A) = \Pvu &\implies \xv^\T = \frac{p^{-1}(\Pvu)}{\mathbb{P}(\neg \S)} \label{eq:subrationalV2VIndifC}
\end{align}

Furthermore, the behavior tuple $x$ satisfying the above conditions is an \emph{essentially unique} equilibrium for $G$, meaning that any equilibrium $x'$ must satisfy the above conditions or have $\Pti(x') = \Pti(x)$. 
\end{lemma}

\begin{proof}
First, we shall show that $x$ is an equilibrium. 
To that end, consider the case where $\Pti(G) > \Pn$. 
By basic algebra, $1-\mathbb{P}(\A) < r\mathbb{P}(\A)$, and by \eqref{eq:NV2VSubrationalAverageCost}, $\Jn(\C, x) < \Jn(\R; x)$. 
Then, note that $\xn^\R = 0$ (respectively $\xn^\C = 1-y$) satisfies \eqref{eq:ActionInUseCond}. 

An identical technique suffices in the cases where $\Pti(G) = \Pn$ or $\Pti(G) < \Pn$. 
Using \eqref{eq:V2VSubrationalAverageCost} in the same manner extends this result to the behavior of of V2V drivers. 
Therefore, $x$ is an equilibrium for $G$. 

It remains to show that any other equilibrium $x'$ is essentially identical to $x$.
Note that unless $\Pti(x') = \Pvs$, $\Pti(x') = \Pn$, or $\Pti(x') = \Pvu$, $x'$ satisfies the above conditions by Lemma~\ref{lemma:subrationalStrictBehavior} and we are finished.
But clearly all equilibria with $\Pti(x) = \Pvs$ have equivalent accident probability (and similarly for $\Pn$ and $\Pvu$), so we are finished. 
%
%
%
%
\end{proof}

Using this result, we partition our parameter space. 
Each of the following sets corresponds to a particular ``family'' of equilibria, restricting the accident probability of games in that set. 
We define these families in equations \eqref{eq:eqFamily1}-\eqref{eq:eqFamily7}. 

\begin{figure*}[ht]
\begin{align}
    E_{1} &:= \{(\beta, y, r) : \Pvu < p(0)\}, \label{eq:eqFamily1} \\
    E_{2} &:= \{(\beta, y, r) : p(0) \le \Pvu \le p(y-(\Pvu(t(y)-f(y))\beta +f(y)\beta)y)\}, \label{eq:eqFamily2} \\
    E_{3} &:= \{(\beta, y, r) : p(y-(\Pvu(t(y)-f(y))\beta +f(y)\beta)y)\} < \Pvu \land \Pn < p(y-(\Pn(t(y)-f(y))\beta +f(y)\beta)y)\}, \label{eq:eqFamily3} \\
    E_{4} &:= \{(\beta, y, r) : p(y-(\Pn(t(y)-f(y))\beta +f(y)\beta)y) \le \Pn \le p(1-(\Pn(t(y)-f(y))\beta +f(y)\beta)y)\}, \label{eq:eqFamily4} \\
    E_{5} &:= \{(\beta, y, r) : p(1-(\Pn(t(y)-f(y))\beta +f(y)\beta)y)\} < \Pn \land \Pvs < p(1-(\Pvs(t(y)-f(y))\beta +f(y)\beta)y)\}, \label{eq:eqFamily5} \\
    E_{6} &:= \{(\beta, y, r) : p(1 - (\Pvs(t(y)-f(y))\beta + f(y)\beta)y) \le \Pvs \le p(1) \}, \label{eq:eqFamily6} \\
    E_{7} &:= \{(\beta, y, r) : p(1) < \Pvs\}. \label{eq:eqFamily7} 
\end{align}

\caption{Equilibrium family partitions for games with endogenous accident probability between non-Bayesian drivers.}
\end{figure*}


These partitions allow us to describe equilibrium accident probability with a finer granularity. 
(This is proved using a technique identical to the one used for Lemma 4.1 in \cite{gould_information_2023}.)

\begin{lemma}
    \label{lemma:subrationalEndoParameterCrashProbMapping}
    For any non-Bayesian game $G = (\beta, y, r)$, $G \in \cup_{i =1}^7 E_i$, and 
    \begin{align}
        G \in E_1 &\implies \Pti(G) = p(0) \\
        G \in E_2 &\implies \Pti(G) = \Pvu \\
        G \in E_3 &\implies \Pn < \Pti(G) < \Pvu \\
        G \in E_4 &\implies \Pti(G) = \Pn \\
        G \in E_5 &\implies \Pvs < \Pti(G) < \Pn \\
        G \in E_6 &\implies \Pti(G) = \Pvs \\
        G \in E_7 &\implies \Pti(G) = p(1)
    \end{align}
\end{lemma}

\begin{proof}
Consider any game $G = (\beta, y, r)$. 
If $G \not \in E_1$ and $G \not \in E_2$, then we must have that 
\[
p((1 - \beta \Pvu (t(y)-f(y)) - \beta f(y))y) < \Pvu.
\]
Similarly, if $G \not \in E_7$ and $G \not \in E_6$, then we must have that 
\[
p((1 - \beta \Pvu (t(y)-f(y)) - \beta f(y))y) < \Pvu.
\]
Therefore, if $G \not \in E_3$ and $G \not \in E_5$,  
\begin{align*}
    &p((1 - \beta \Pn (t(y)-f(y)) - \beta f(y))y) \le \Pn \text{ and} \\ 
    &\Pn \le p(1 - (\beta \Pn (t(y)-f(y)) + \beta f(y))y).
\end{align*}
But this implies that $G \in E_4$. 
Therefore, $G \in \bigcup_{i=1}^7 E_i$, as desired. 

We shall now prove that $G \in E_3 \implies \Pn < \Pti(G) < \Pvu$. 
For brevity, the other claims are omitted, but can be proved with an identical method. 
Let $G \in E_3$, then
\begin{align*}
    &p((1 - \beta \Pvu (t(y)-f(y)) - \beta f(y))y) < \Pvu \text{ and } \\ 
    &\Pn < p((1 - \beta \Pn (t(y)-f(y)) - \beta f(y))y).
\end{align*}
If $\beta t(y) = 0$, the above conditions give a contradiction, so $\beta t(y) > 0$. 

Assume by contradiction that $\Pti(G) \le \Pn$, and let $x$ be an equilibrium of $G$. 
Then, since $\beta t(y) > 0$, $\Pti(G) < \Pvu$, so by \eqref{eq:V2VTrustCareful} and \eqref{eq:ActionInUseCond}, $\xv^\C = 0$. 
This gives that $\xv^\R + \xv^\T = \xv = y$, meaning $\xv^\R + \mathbb{P}(\neg \S) \ge \mathbb{P}(\neg \S)y$.
Further, it is always true that $\xn^\R \ge 0$. 
Therefore, By \eqref{eq:subrationalPRecursion} and the fact that $p$ is increasing,
\begin{align*}
    \Pti(G) &= p(\xn^\R + \xv^\R + \mathbb{P}(\neg \S)\xv^\T) \\ 
    &\ge p(\mathbb{P}(\neg \S)) \\ 
    &= p(y-(\Pti(G)(t(y)-f(y))\beta + f(y)\beta)y)
\end{align*}

Then, starting with our contradiction hypothesis, we perform algebraic operations to take $\Pti(G)$ ``up'' one level of its recursive definition in \eqref{eq:subrationalPRecursion}.
This gives that 
\[
\begin{split}
    y - (\Pti(G)(t(y)-f(y))\beta + f(y)\beta)y \ge \\ 
    y - (\Pn(t(y)-f(y))\beta + f(y)\beta)y,
\end{split}
\]
and since $p$ is increasing, 
\[
\begin{split}
    p(y - (\Pti(G)(t(y)-f(y))\beta + f(y)\beta)y) \ge \\ 
    p(y - (\Pn(t(y)-f(y))\beta + f(y)\beta)y).
\end{split}
\]

But then we have the following chain of inequalities
\begin{align*}
    \Pn &< p(y - (\Pn(t(y)-f(y))\beta + f(y)\beta)y) \\ 
    &\le p(y - (\Pti(G)(t(y)-f(y))\beta + f(y)\beta)y) \\ 
    &\le \Pti(G) \le \Pn,
\end{align*}
which is a contradiction. 
This completes the proof. 
\end{proof}

We are now equipped to describe the essentially unique equilibrium of any game $G$;  Lemma~\ref{lemma:subrationalEndoParameterCrashProbMapping} gives a restriction on equilibrium accident probability from game parameters, and Lemmas~\ref{lemma:subrationalStrictBehavior} and \ref{lemma:subrationalEndoEqUniqueness} yield driver behavior under this restriction.
This behavior is sufficient to show that the accident probability induced by non-Bayesian drivers is the same as that of Bayesian drivers. 

\begin{lemma}
    \label{lemma:endoCPEquality}
    For any game $G = (\beta, y, r)$ with endogenous accident probability, 
    \begin{equation}
        \label{eq:endoCPEqualityProof}
        \Ptb(G) = \Pti(G). 
    \end{equation}
\end{lemma}

\begin{proof}
    Lemma~\ref{lemma:subrationalEndoParameterCrashProbMapping} shows that every game $G$ belongs to one of the equilibrium families defined by \eqref{eq:eqFamily1}-\eqref{eq:eqFamily7}, meaning we can show this simply using cases. 

    Equilibrium accident probability is restricted as a function of game parameters by Lemma~\ref{lemma:subrationalEndoParameterCrashProbMapping} for non-Bayesian agents and by \cite[Lemma 4.1]{gould_information_2023} for Bayesian agents. 
    Accident probability in both cases is restricted to a single value and we are immediately finished unless $G \in E_3$ or $G \in E_5$. 
    In the first case, note that Bayesian agents choose $\xn^\R = 0$, $\xvs^\R = 0$, and $\xvu^\R = \mathbb{P}(\neg \S)y$ as a consequence of \cite[Lemma 4.2]{gould_information_2023}.
    Similarly, non-Bayesian agents choose $\xn^\R = 0$, $\xv^\R = 0$, and $\xv^\T = y$ by Lemma~\ref{lemma:subrationalEndoParameterCrashProbMapping}. 
    Then, by \eqref{eq:rationalPRecursion} and \eqref{eq:subrationalPRecursion},
    \[
        \Ptb(G) = p(\mathbb{P}(\neg \S)y) = \Pti(G),
    \]
    which is the desired result. 

    This can be shown in a similar manner if $G \in E_5$. completing the proof. 
\end{proof}

\addtolength{\textheight}{-13mm}

\begin{lemma}
    \label{lemma:endoSCEquality}
    For any game $G = (\beta, y, r)$ with endogenous accident probability, 
    \begin{equation}
        \label{eq:endoSCEqualityProof}
        \SCtb(G) = \SCti(G). 
    \end{equation}
\end{lemma}

\begin{proof}
    Again by Lemma~\ref{lemma:subrationalEndoParameterCrashProbMapping}, we show this by cases.
    Note that for Bayesian agents, \eqref{eq:socialCostDef} simplifies to 
    \[
    \begin{split}
        \SCtb(G) &= \Jn (\C; x) \xn^\C + \Jn (\R; x) \xn^\R \\
        &+ \Jvu (\C; x) \xvu^\C + \Jvu (\R; x) \xvu^\R \\
        &+ \Jvs (\C; x) \xvs^\C + \Jvs (\R; x) \xvs^\R,
    \end{split}
    \label{eq:rationalSCImpl}
    \]
    and for non-Bayesian agents it simplifies to
    \[
    \begin{split}
        \SCti(G) &= \Jn(\C; x) \xn^\C + \Jn(\R; x)\xn^\R \\ 
        &+ \Jv(\C; x) \xv^\C + \Jv(\R; x) \xv^\R + \Jv(\T;x) + \xv^\T. 
    \end{split}
    \label{eq:subrationalSCImpl}
    \]

    Consider the case where $G \in E_2$. 
    Bayesian agents choose behaviors $\xn^\R = 0$, $\xvs^\R = 0$, and $\xvu^\R = p^{-1}(\Pvs)$ (as a consequence of \cite[Lemma 4.2]{gould_information_2023}), and non-Bayesian agents choose $\xn^\R = 0$, $\xv^\R = 0$, and $\xv^\T = y$ by Lemma~\ref{lemma:subrationalEndoParameterCrashProbMapping}. 
    Then, both of the above simplify to give 
    \[
        \SCtb(G) = \frac{r(1-\beta t(y)}{1-\beta f(y) +r(1-\beta t(y))} = \SCti(G),
    \]
    which is the desired result. 

    A series of (tedious) algebraic calculations give the same result in each other case, completing the proof. 
\end{proof}

\section{Conclusion}
\label{sec:conclusion}
This work considered a class of models describing how human agents respond to road hazard information sharing. 
We discussed two independent characteristics of these models, and their effects on equilibrium behavior of drivers. 
We first showed that games of this kind with incomplete information can be equivalently interpreted as imperfect information games, removing an assumption on human rationality. 
We used this fact to prove our main result, that models with an endogenous accident probability can describe scenarios that those with an exogenous accident probability cannot; namely, that social cost can be increasing with information quality. 
Future work could consider additional alternative descriptions of human behavior, and possibly use a heterogeneous population  of these different descriptions. 

\bibliographystyle{ieeetr}
\bibliography{ref/ref_gould_V2V,ref/library}

\end{document}